# Synthesis of ultrahigh-quality monolayer molybdenum disulfide through *in-situ* defect healing with thiol molecules

Simin Feng[1], Junyang Tan[1], Shilong Zhao[1], Shuqing Zhang[1], Usman Khan[1], Lei Tang[1], Xiaolong Zou[1], Junhao Lin[2], Hui-Ming Cheng[1,3,*], Bilu Liu[1,*]

[1] Shenzhen Geim Graphene Center, Tsinghua-Berkeley Shenzhen Institute and Tsinghua Shenzhen International Graduate School, Tsinghua University, Shenzhen, 518055, P. R. China

[2] Shenzhen Key Laboratory of for Advanced Quantum Functional Materials and Devices, Department of Physics, Southern University of Science and Technology, Shenzhen, 518055, P. R. China

[3] Shenyang National Laboratory for Materials Sciences, Institute of Metal Research, Chinese Academy of Sciences, Shenyang, 110016, P. R. China

Correspondence should be addressed to B.L. (bilu.liu@sz.tsinghua.edu.cn), or H.M.C. (hmcheng@sz.tsinghua.edu.cn)

**ABSTRACT.**

Monolayer transition metal dichalcogenides (TMDCs) are two-dimensional (2D) materials with many potential applications. Chemical vapour deposition (CVD) is a promising method to synthesize these materials. However, CVD-grown materials generally have poorer quality than mechanically exfoliated ones and contain more defects due to the difficulties in controlling precursors' distribution and concentration during growth where solid precursors are used. Here, we propose to use thiol as a liquid precursor for CVD growth of high quality and uniform 2D $MoS_2$. Atomic-resolved structure characterizations indicate that the concentration of sulfur vacancies in the $MoS_2$ grown from thiol is the lowest among all reported CVD samples. Low temperature spectroscopic characterization further reveals the ultrahigh optical quality of the grown $MoS_2$. Density functional theory simulations indicate that thiol molecules could interact with sulfur vacancies in $MoS_2$ and repair these defects during the growth of $MoS_2$, resulting in high quality $MoS_2$. This work provides a facile and controllable method for the growth of high-quality 2D materials with ultralow sulfur vacancies and high optical quality, which will benefit their optoelectronic applications.

**KEYWORDS**

Two-dimensional materials, molybdenum disulfide, thiol molecule, chemical vapor deposition, defect, low temperature photoluminescence

**INTRODUCTION.**

Mono and few-layer semiconducting transition metal dichalcogenides (TMDCs) in the form of $MX_2$ (M = Mo, W; X=S, Se), are two-dimensional (2D) materials with interesting properties that complement those of graphene and hexagonal boron nitride (h-BN). [1] The researches and applications of TMDCs greatly rely on efficient methods to obtain monolayers from their bulk crystals and, in addition, on the direct growth of their flakes. [2] Among all these methods, chemical vapor deposition (CVD) has been considered as a promising method as it provides a relatively scalable and controllable way for growing TMDCs.[3-12] Current CVD method usually uses solid (such as metal oxides and sulfur powder) as precursors for the growth of TMDCs. However, due to the significant differences of vapor pressure between sulfur and metal oxides, the reaction usually needs precise control of temperature and pressure in order to achieve stable supply of precursors, leading to poor controllability and reproducibility. To solve this problem, Ling *et al.* and Li *et al.* have used salt additives during the growth process.[13, 14] These additives form intermediate volatile compounds with metal oxides, so as to promote the reaction and expand the thermodynamic window of chemical reaction, which is the key factor to enable the batch production of 6-inch growth of monolayer $MoS_2$[15] and the exploration of the entire TMDC library,[16] proving the universality of this method in growing TMDCs. Although the syntheses of TMDCs has made great progress, the use of solid-state precursors still has the problem of uncontrollable reaction rate. Unlike graphene growth with a constant supply of gas precursors, for TMDCs, the volatilization rate of solid powder is very sensitive to temperature and pressure, so that the quality of the materials obtained is largely determined by the position of substrates. For example, Wang *et al.* have studied the relation between the morphology of as-grown $MoS_2$ and the position of the substrate. They found that the evaporation of solid precursors forms a density gradient in the furnace, leading to high dependence of $MoS_2$ morphology on the position of the substrate.[17, 18] Furthermore, it has been proved that adding salt during growth process introduces impurities, thus decreases the quality and influences the stability of the sample.[19] Therefore, it is crucial to develop a new method to obtain uniformly distributed 2D $MoS_2$ with high quality.

Here, we developed a liquid-precursor CVD method (LCVD) and achieved the growth of MoS$_2$ monolayers with ultrahigh quality. By bubbling sulfur-containing thiol (dodecyl mercaptan, C$_{12}$H$_{25}$SH) solution into a furnace, the precursor concentration during growth can be precisely controlled and consequently homogeneously-distributed MoS$_2$ are grown on substrates. Systematic microscopic and spectroscopic characterizations confirm the good uniformity and quality of MoS$_2$ monolayers, having the lowest density of sulfur vacancies and highest optical quality among the CVD samples reported previously and very close to the exfoliated MoS$_2$ by using a scotch tape. Theoretical simulations show that thiol molecules can effectively repair the sulfur vacancies in MoS$_2$ during the CVD process, resulting in high quality. These results demonstrate that LCVD is an efficient method to synthesize monolayer MoS$_2$ with ultralow sulfur vacancies and excellent optical properties, which is significantly important for MoS$_2$ fabrication and photonic applications.

**RESULTS AND DISCUSSION.**

The growth process is illustrated in **Figure 1**. The upper panel in **Figure 1a** shows a typical solid-precursor CVD (SCVD) system for the growth of 2D TMDCs where solid powders (e.g., MoO$_3$ and S powders) are used as precursors. This traditional system suffers from the poor controllability of precursors and the non-uniform growth of samples due to the gradient diffusion of precursor concentrations which depend on the distance between substrate and solid sources. In addition, as reaction continues, the amount of precursors decreases, leading to insufficient supply of precursors and resulting in the non-uniform growth of MoS$_2$ with many defects (**Figure 1b-c**).[20] Inspired by the growth of large-area, uniform and high quality graphene domains and films, we developed the LCVD method and the schemes are shown in the bottom panel of **Figure 1a** and **Figure S1a**. In short, a SiO$_2$/Si wafer was spin-coated with Na$_2$MoO$_4$ aqueous solution and put in the center of a furnace as a growth substrate. Thiol was put in a container upstream and was bubbled into the furnace by Ar gas as sulfur precursor. The growth was conducted at 850 °C in Ar (see experimental details in the **Experimental Section**). Unlike SCVD method which usually shows a position-dependent distribution of MoS$_2$ flakes with abundant defects, constant supply and effective control of precursors will lead to ultrahigh quality MoS$_2$ growth (**Figure 1d)**). Furthermore, we observe from optical images that LCVD method results in a uniform growth of monolayer MoS$_2$ across the substrate (**Figure 1e and Figure S1b**).

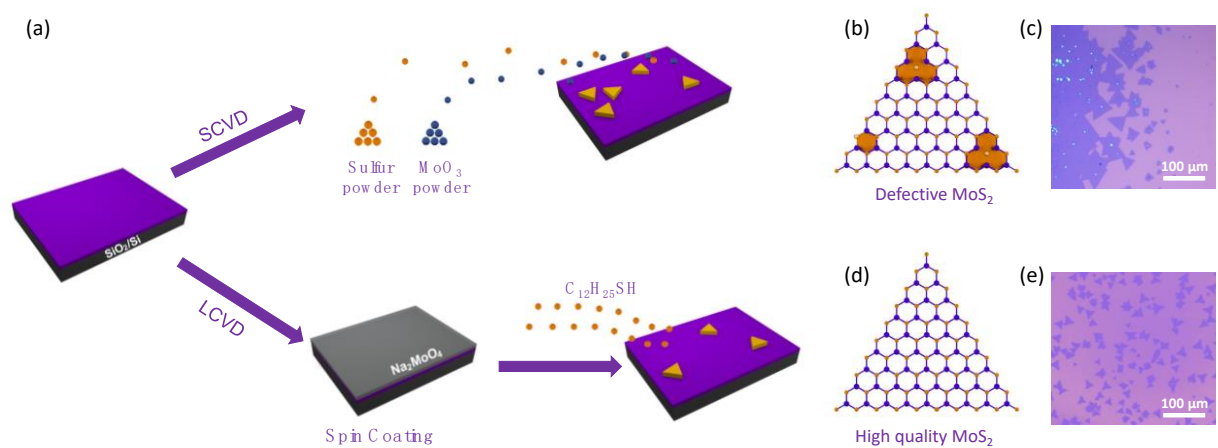

**Figure 1. LCVD growth of MoS$_2$ in comparison with traditional CVD.** (a) Schematic of LCVD growth process compared with SCVD. (b) Illustration of the defective MoS$_2$ grown by SCVD. (c) Optical microscopic image of MoS$_2$ grown by SCVD, showing a non-uniform growth pattern across the substrate. (d) Illustration of the high quality MoS$_2$ grown by LCVD. (e) Optical microscopic image of MoS$_2$ grown by LCVD, showing a uniform growth of monolayer MoS$_2$ across the substrate.

Raman and PL spectroscopy were used as non-destructive techniques to characterize MoS$_2$ grown using thiol. **Figure 2a** shows a representative PL spectrum obtained from monolayer

MoS$_2$ grown by LCVD, which exhibits a sharp and narrow peak located around 1.84 eV (A exciton) and a small shoulder located around 2.0 eV (B exciton), corresponding to the direct bandgap transition of monolayer MoS$_2$. The inset of **Figure 2a** shows a typical Raman spectrum of MoS$_2$ which exhibits the characteristic in-plane E' phonon mode located at 385 cm$^{-1}$ and out-of-plane A$_1$' mode at 404 cm$^{-1}$. These modes are analog to the E$_{2g}$ and A$_{1g}$ peak observed from bulk MoS$_2$, but with different names due to the different group symmetry between bulk (D$_{6h}$ group) and monolayer MoS$_2$ (D$_{3h}$ group).[21] Besides these, light reflectivity measurement constitutes another powerful tool to study and identify TMDCs. We measured the reflectance contrast spectra (R$_{sample}$-R$_{sub}$)/R$_{sub}$ of monolayer MoS$_2$, where R$_{sample}$ is the reflectivity of monolayer MoS$_2$ on a 285 nm thick SiO$_2$/Si substrate and R$_{sub}$ is the reflectivity of the substrate. **Figure S2** shows the optical path for the measurements and **Figure 2b** shows the reflectance contrast measurement in the energy range of 1.6 - 2.4 eV. Two predominate peaks at ~1.84 eV and ~1.99 eV can be clearly observed, which matches well with our PL measurements. The broad peak at ~2.22 eV is related to the slowly varied background caused by the 285 nm thick SiO$_2$/Si substrate. The above spectroscopy results show the successful growth of monolayer MoS$_2$ by LCVD.

Next we focus on the uniformity of each flake. We randomly chose 20 different flakes and analyzed their Raman and PL spectra (**Figure S3**). All the flakes show frequency differences between A$_1$' and E' Raman peaks in a narrow range of 18.3 – 19.0 cm$^{-1}$ and the peak intensity ratio between Si and MoS$_2$ A$_1$' Raman peaks in a range of 4.0 - 4.5 (**Figure 2c**). The Raman frequency spacing between A$_1$' and E' peaks could be an indicator of MoS$_2$ number of layers and the results obtained here are similar to monolayer MoS$_2$ reported previously,[22] indicating the growth of uniform monolayer $MoS_2$ across the substrate. In addition to Raman results, the PL peak position and intensity plotted as a function of these 20 flakes show uniform monolayer $MoS_2$ domains as well (**Figure 2d**). Similarly, **Figure 2e** statistically summarizes the peak positions of A and B excitons of 10 different $MoS_2$ flakes and all these flakes exhibit consistent reflectivity peak at around 1.995 eV, showing good uniformity of these flakes. In terms of the uniformity of a single $MoS_2$ flake, we performed Raman and PL mapping on their characteristic peaks, which show uniform intensity over the individual flake (**Figures 2f–2h**). All these spectroscopy results confirm the high uniformity and crystalline quality of $MoS_2$ flakes grown by LCVD method.

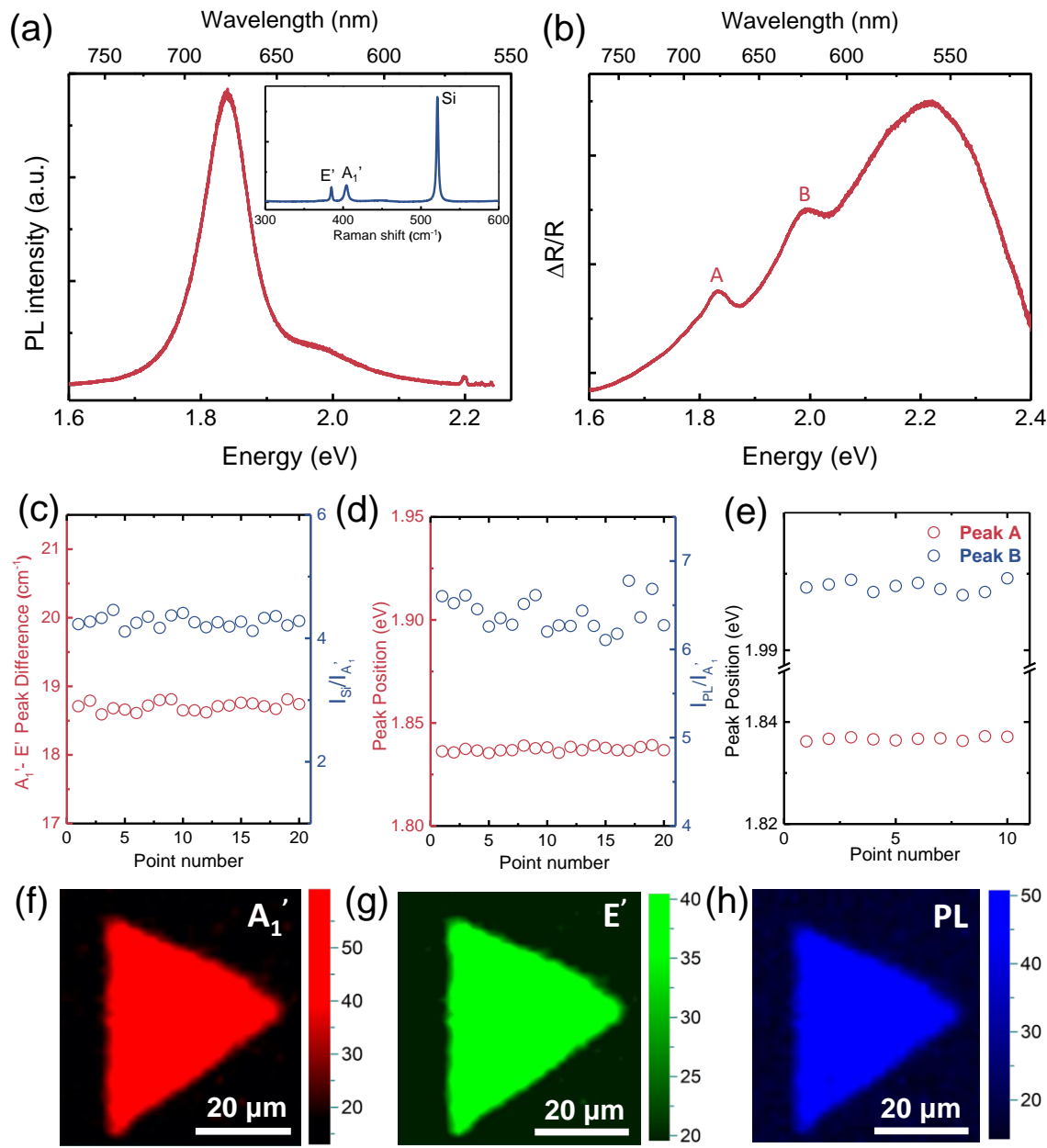

**Figure 2. High quality and uniformity of LCVD-grown monolayer MoS₂ flakes.** (a) Typical PL spectrum of LCVD MoS$_2$, showing a sharp and narrow peak at 1.84 eV. Inset shows a typical Raman spectrum of LCVD MoS$_2$, showing A$_1$' and E' MoS$_2$ Raman fingerprints and Si Raman peaks. (b) Typical differential reflectivity of LCVD MoS$_2$ flakes. A and B exciton peaks are labelled here. (c-d) Statistic analysis of (c) LCVD MoS$_2$ A$_1$' Raman peak intensity (left column) and peak position spacing between A$_1$' and E' peak (right column) and (d) LCVD MoS$_2$ PL peak position (left column) and peak intensity (right column), measured on 20 MoS$_2$ flakes randomly chosen from the substrate. (e) Differential reflectivity statistical analysis of MoS$_2$ A and B exciton peak position measured on 10 MoS$_2$ flakes randomly chosen from the substrate. (f-h) Raman and PL mapping of one LCVD MoS$_2$ flake, showing uniform Raman and PL intensity across the sample.

We then focus on the microstructure of the samples. A low magnified scanning transmission electron microscope (STEM) image (**Figure 3a**) shows the morphology of intact triangle-shaped $MoS_2$ after the transfer process, with only one set of hexagonal diffraction pattern as shown in the inset, confirming its hexagonal symmetry and single-crystal nature. High resolution TEM and its inverse fast Fourier transform (IFFT) images (**Figure S4a**) reveal the honeycomb-like structure of the $MoS_2$ lattice with a lattice constant of 0.28 nm for (010) plane. Energy-dispersive spectroscopy (EDS) mapping shows the presence of Mo and S elements in the triangular region (**Figure S4b-S4e**) without obvious impurities.

To further elucidate the crystalline quality of LCVD $MoS_2$ at atomic scale, aberration corrected high-angle annular dark field scanning TEM (HAADF-STEM) imaging was performed to study the intrinsic defects in the as-grown $MoS_2$ (**Figure 3b**). Due to the Z-contrast nature of HAADF imaging, the mono S vacancy (SV) (red arrow in top panel of **Figure 3c**) could be distinguished from brighter $S_2$ column, which is in good arrangement with STEM simulation (middle panel in **Figure 3c**), and also supported by intensity profile analysis (bottom panel in **Figure 3c** and **Figure S5**). For quantitative analysis of SV concentration, an intensity histogram of all sites from Figure 3b is shown in **Figure 3d**. It could be observed that there is only one peak for Mo sites, while for the case of S sites, besides the main peak representing the two overlapped S atoms, there is also a small trail with lower intensity, *i.e.*, the sites of mono SV. Based on the HAADF intensity of each site, an atom-by-atom model is built for all sites in **Figure 3e** with mono SV marked by blue circles in Figure 3b correspondingly. Consequently, the density of SV was calculated to be $0.32/nm^2$, which is consistent across the whole samples with a statistical SV density of around $0.34/nm^2$ (from several randomly picked areas in LCVD sample, **Figure S6**). Noteworthy, we have compared the SV density of LCVD MoS$_2$ with MoS$_2$ grown by conventional CVD method with solid precursors and the exfoliated MoS$_2$ (**Figure 3f**). The SV density of monolayer MoS$_2$ grown by LCVD is the lowest among all reported CVD grown samples, [8, 23-27] and close to the exfoliated one (0.15/nm$^2$, **Figure S7**). Besides, it is worth mentioning that all the point defects are mono SV, which is different from previous reports of SCVD-grown MoS$_2$ where di-sulfur vacancies are usually observed. [8, 25, 27]

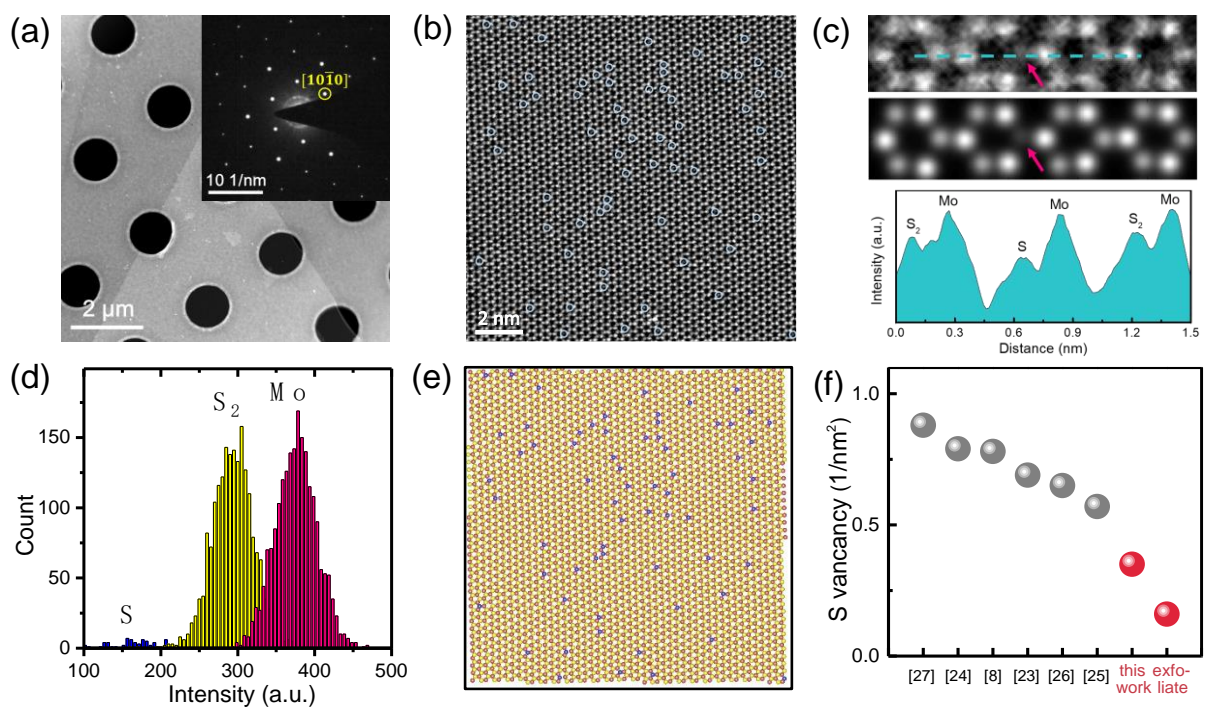

**Figure 3. Ultralow sulfur vacancies of LCVD-grown MoS$_2$.** (a) Low magnification HAADF image of LCVD MoS$_2$. The inset shows the selected area electron diffraction (SAED) of LCVD MoS$_2$. (b) HAADF-STEM image of LCVD MoS$_2$ flake with an area of ~ 191 nm$^2$, and the mono-sulfur vacancies are all marked with blue circles. (c) Experimental (top) and simulated (middle) HAADF-STEM images of monolayer MoS$_2$ with a mono-S vacancy marked by red arrow and its corresponding intensity line profile (bottom) taken from the annotated line of experimental results. (d) Histogram of intensity based on results from (c). Here "S$_2$" means double S atoms and "S" means single S atom (mono-S vacancy). (e) False color elemental mapping of Mo and S sites based on results from (b). Mo and double S atoms are marked with red and yellow while single S atom is marked with blue. (f) Comparison of SV intensity of LCVD MoS$_2$ with other MoS$_2$ samples reported from previous literature and exfoliated MoS$_2$ sample.

To further investigate the optical quality of our samples, we have performed PL measurements at low temperature, because PL spectra of exciton emission is sensitive to defect density and could be another crucial indicator of material quality.[28-30] **Figure 4a** shows a sequence of PL spectra at temperatures ranging from 300 K to 80 K with a 20 K temperature interval and the corresponding color contour mapping image is shown in **Figure 4b**. At 80 K, the full spectra could be fitted into four peaks, as shown in **Figure 4c**. The most dominant feature contains two peaks, labelled as neutral A excitons (~ 1.91 eV) and negatively charged trions (~ 1.89 eV). Trion is formed by trapping an excess electron to neutral exciton with a binding energy of ~ 20 meV.[31] A third peak (B excitons) could be observed at ~ 2.08 eV. Both A and B exciton peaks are associated with optical transitions at K point of the first Brillouin zone.[1, 32, 33] The energy differences between these two peaks arise from the spin-orbital splitting of the valence band energy, as depicted in the inset in **Figure 4a**. In addition to these three peaks, a lower energy peak resulted from the radiative recombination of deep level defect-bounded excitons ($X_B$) appears at low temperature.[20, 34] These bound excitons can arise from the presence of defects or impurities such as sulfur vacancies in monolayer $MoS_2$.[35] The reason that this peak is not observable at room temperature is due to the low binding energy of free excitons bound to the defect states, which are easy to dissociate when the temperature increases. The fact that the peak intensity of this defect related peak is significantly smaller than A exciton peak is a clear indicator of a low defect density sample, in agreement with our HAADF-STEM observations.

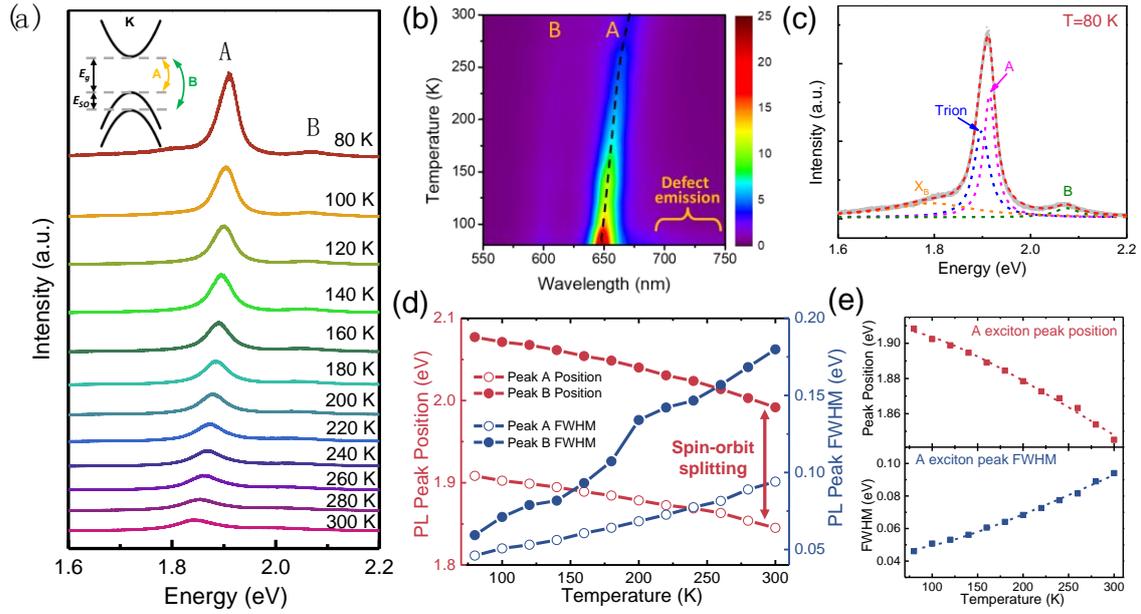

**Figure 4. Low temperature PL measurements of LCVD-grown monolayer MoS$_2$.** (a) Temperature dependent PL spectra of LCVD MoS$_2$, A and B denote excitons, and the inset shows the band structure near the K point of the Brillouin zone indicating the direct band transition (E$_g$) and spin-orbit splitting (E$_{SO}$). (b) A color contour plot of the temperature dependence of the photoluminescence intensity. The dotted line tracks the evolution of the peak position of the neutral A exciton. Wavelength ranges from 680 nm to 750 nm are related to defect-bounded emission as labelled by "defect emission". (c) Curve fitting of low temperature PL spectrum into 4 peaks, labelled as neutral A exciton (A), neutral B exciton (B), negatively charged A exciton (trion) and defect bounded emission (X$_B$). (d) Summary of temperature evolution of peak position and linewidth for A and B exciton peaks. (e) A exciton peak position and linewidth versus temperature and its best fit to the semi-empirical model (dashed lines).

To gain further insights into the main exciton transition, the evolution of the peak position and linewidth as a function of temperature for A and B excitons are analyzed and plotted in **Figure 4d**. We observe that both A and B exciton peaks become narrow and blue-shifted when the temperature decreases due to widened bandgap from lattice shrinkage and less thermal relaxation at low temperature. The emission energy of the neutral exciton A as a function of temperature is described by the O'Donnell model, derived from the thermodynamics of electron-hole pairs in semiconductor and is written as: [36, 37]

$$E_{X^0}(T) = E_{X^0}(0) - S\langle\hbar\omega\rangle\left[\coth\left(\frac{\langle\hbar\omega\rangle}{2k_BT}\right) - 1\right] \quad (1)$$

Where $E_{X^0}(0)$, S, $k_B$, and $\langle\hbar\omega\rangle$ are the neutral exciton energy at zero temperature, a dimensionless electron-phonon coupling constant, Bolzmann's constant, and an average phonon energy, respectively. The fitted curve (red dashed line shown in **Figure 4e**) yields $E_{X^0}(0) = 1.961 \pm 0.011$ eV, $S = 2.02 \pm 0.31$ and $\langle\hbar\omega\rangle = 25.27 \pm 4.59\ meV$. The coupling constant and the average phonon energy are similar to those obtained in the literature.[24, 33, 38-41] The evolution of the linewidth as a function of temperature is shown in **Figure 4e** and can be phenomenologically approximated by a phonon induced broadening effect (blue dashed line):[40, 42]

$$\gamma = \gamma_0 + c_1 T + \frac{c_2}{e^{\langle\hbar\omega\rangle/k_BT} - 1} \quad (2)$$

where $\gamma_0 = 42.6 \pm 1.3\ meV$ is the temperature independent inhomogeneous broadening, $c_1 = 19 \pm 5\ \mu eV/K$ describes the linear increase due to acoustic Γ phonons, $c_2 = 87 \pm 18\ meV$ is a measure of the strength of exciton-phonon coupling, and $\langle\hbar\omega\rangle = 25.3\ meV$ is the averaged energy of the relevant phonon, which is obtained by fitting the A exciton energy shift with Equation (1) for consistency.

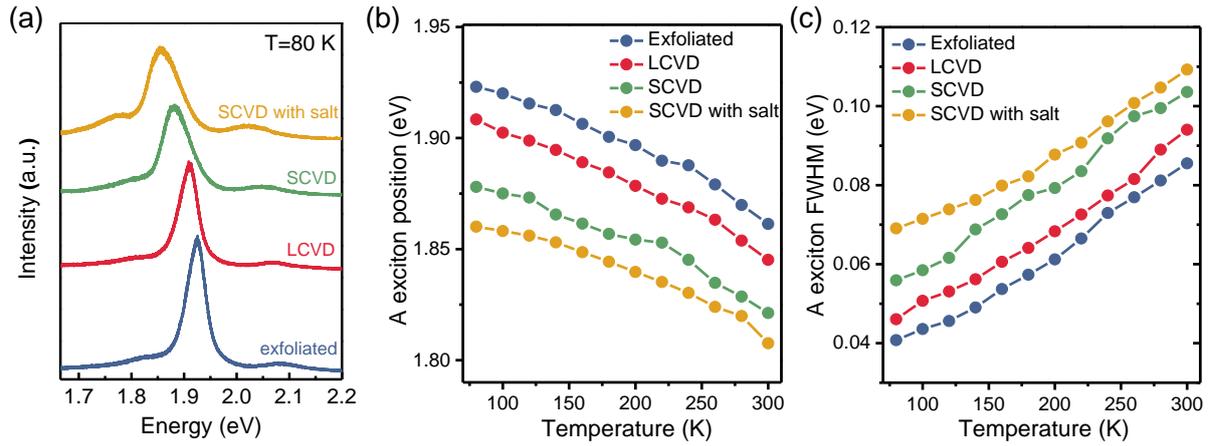

**Figure 5. Comparison of temperature dependent A exciton peak position and linewidth of LCVD MoS$_2$ with those of exfoliated and SCVD MoS$_2$.** (a) Temperature dependent PL spectra of MoS$_2$ grown by different methods. (b) and (c) Comparison of temperature dependent A exciton peak position and linewidth of LCVD MoS$_2$ with those of exfoliated and SCVD MoS$_2$.

To further verify the quality of our sample, we compare the low temperature PL spectra of LCVD MoS$_2$ with those of exfoliated monolayer MoS$_2$ and SCVD MoS$_2$ grown with and without adding salt. The typical PL spectra of all the samples at 80 K are shown in **Figure 5a**. The evolution of peak position and linewidth with respect to temperature for all four different samples are plotted in **Figure 5b and 5c** (also in **Figure S8**). It should be noted in **Figure 5b** that the peak position of neutral A exciton for LCVD MoS$_2$ is closest to that of exfoliated samples and only differs by ~ 15 meV, while the peak position for SCVD MoS$_2$ grown with salt red shifted by 70 meV. This shift has been ascribed to bandgap renormalization modulated by defect density [43] and the almost unchanged exciton peak position for LCVD MoS$_2$ indicates the least defect density. An additional important parameter indicative of sample quality in TMDCs is the PL linewidth at low temperature, which reflects intrinsic contributions from the radiative lifetime of exciton-phonon coupling, as well as extrinsic inhomogeneous broadening effect from factors such as defects in the material and substrate-induced carrier doping. We observe that the exfoliated MoS$_2$ exhibits a full width at half maximum (FWHM) of 40 meV at 80 K (**Figure 5c**). The FWHM of LCVD MoS$_2$ is almost identical (44 meV) to that of the exfoliated MoS$_2$ and is much narrower than that of SCVD samples, indicating the comparable defect density of LCVD MoS$_2$ with that of the exfoliated material. All these results have proven that LCVD grown MoS$_2$ and exfoliated material possess comparable optical quality.

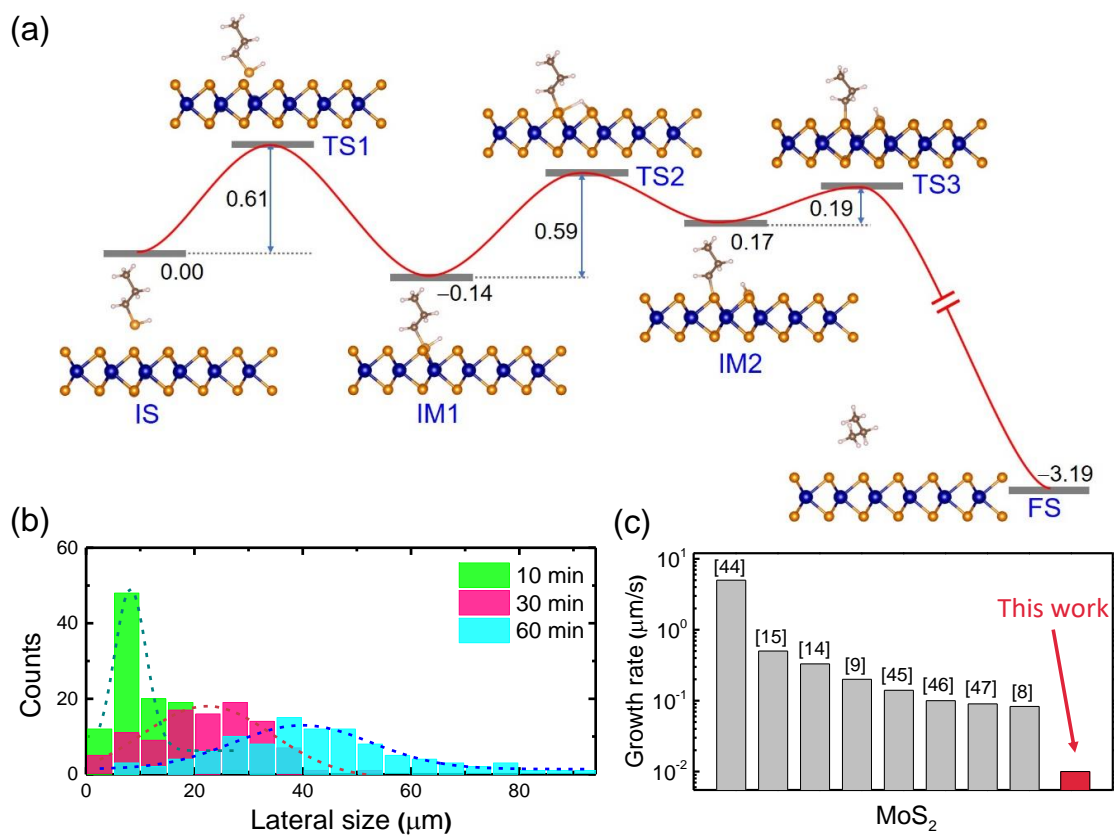

**Figure 6. *In-situ* SV repairing process by thiol molecules during LCVD.** (a) Kinetics and transient states of the reaction between a mono-S vacancy and propanethiol molecule. Numbers represent relative energies and energy barriers in eV. Repairing pathways start from the initial state (IS) to the final state (FS) via two intermediate states (IM1, IM2) and three transition states (TS1, TS2, TS3). (b) Statistical summary of the lateral sizes of LCVD grown MoS$_2$ with different growth time. A growth rate of 0.01 μm/s was extracted from the fitting. (c) Comparison of the growth rates of monolayer MoS$_2$ grown on SiO$_2$/Si substrates by LCVD and others reported in literature. [8, 9, 14, 15, 44-47]

To shed light on the growth mechanism of monolayer MoS$_2$ grown by LCVD method with thiol precursor and explain why the sample possesses much less SV than those by conventional methods, we carried out density functional theory (DFT) calculations to simulate the reaction kinetics between MoS$_2$ with SV (MoS$_2$-SV) and thiol molecules. The reaction between MoS$_2$-SV and thiol molecules can be described as CH$_3$(CH$_2$)$_n$CH$_2$SH+MoS$_2$-SV → MoS$_2$ + CH$_3$(CH$_2$)$_n$CH$_3$. For simplicity, the case of propanethiol (CH$_3$CH$_2$CH$_2$SH, n=1) is selected as an example, and the reaction process comprises three steps (**Figure 6a**). The first step is the transition from physical adsorption (IS) to chemical adsorption (IM1), where a thiol molecule undergoes an energy barrier of ~0.61 eV (TS1) to approach and connect to SV. The second step is the S-H bond breaking of the thiol molecule (from IM1 to IM2), which experiences an energy barrier of 0.59 eV (TS2). The final step involves the breaking of S-C bond (from IM2 to FS) and the bond formation of C-H, with a barrier as low as 0.19 eV. Once the S-C bond breaks, alkanes with saturated hydrocarbons are easy to form, leading to a sharp decrease in energy (~ 3.19 eV), which is much larger than the energy barriers in reactions. Similar transition states could be observed for the case of butanethiol (CH$_3$(CH$_2$)$_2$CH$_2$SH, n=2), where the energy barriers in three processes are less than 0.6 eV and the whole reaction process is exothermic with an energy change of ~ 3 eV (**Figure S9**). The relative low energy barriers facilitate the saturation of SV in MoS$_2$ and we expect all other alkanethiol molecules, including dodecyl mercaptan, would have similar effects. This is in good agreement with previous results where researchers have post-treated defective MoS$_2$ with thiol molecules to repair the SV and improve the interfacial properties. [26, 48, 49] To better elucidate this process, we have performed control experiments with various growth times. Lateral sizes of 100 MoS$_2$ triangles for each reaction time was measured and are summarized in **Figure 6b**, and a growth rate of ~ 0.01 μm/s was extracted from these results. We compare this growth rate with previous published results of

MoS$_2$ and find that it is at least one order of magnitude smaller. [8, 9, 14, 15, 44-47] This could be explained that dodecyl mercaptan molecules may frequently absorb onto and repair the SV sites in MoS$_2$, leading to a slower supply of sulfur atoms and accordingly the slower growth rate of MoS$_2$. Therefore, we speculate that dodecyl mercaptan would *in-situ* repair SVs while serving as sulfur precursors for the growth of monolayer MoS$_2$, resulting lower SV than those by other conventional CVD methods.

**CONCLUSIONS.**

We have developed a liquid-precursor CVD system and synthesized high quality monolayer MoS$_2$. In this LCVD method, precursor distribution and concentration could be precisely controlled and ultrahigh quality monolayer MoS$_2$ could be achieved. Microscopic and spectroscopic characterizations indicate that the sulfur vacancies in the LCVD MoS$_2$ is the lowest among all CVD samples and is close to those of exfoliated MoS$_2$ flakes. The ultrahigh quality of the the LCVD MoS$_2$ is further proved by low temperature PL measurements, where LCVD samples exhibit a much narrower and sharper PL peak. Theoretical simulations indicate that thiol molecules would chemically absorb onto SV in MoS$_2$ and repair these defects during growth. These results demonstrate that this facile and controllable method can obtain TMDCs with ultralow sulfur vacancies and ultrahigh optical quality, which will facilitate their photonic and optoelectronic applications.

**EXPERIMENTAL SECTION.**

**Materials and chemicals.** Sodium molybdate dehydrate ($Na_2MoO_4 \cdot 2H_2O$, 99.5%, Aladdin Industrial Corporation, China), dodecyl mercaptan ($C_{12}H_{25}SH$, 98%, Aladdin Industrial Corporation, China), molybdenum (VI) oxide powder ($MoO_3$, 99.99%, Alfa Aesar, USA), sulfur powder (S, 99.95%, Sigma-Aldrich, USA), potassium iodide powder (KI, 99%, Alfa Aesar, USA), $SiO_2$/Si substrate (300-nm-thick thermally-grown oxide layer, Hefei Kejing Materials Technology Co., China), acetone, ethanol and isopropyl alcohol (all were Analytical Reagent, Shanghai Macklin Biochemical Co., Ltd., China) were used as received. Bulk $MoS_2$ crystals were used for exfoliation to prepare 2D $MoS_2$ flakes (2D semiconductors, USA).

**LCVD growth of $MoS_2$.** Growth experiments were conducted in a lab-made two-zone CVD furnace equipped with a 1-inch diameter quartz tube (TF55035C-1, Lindberg/Blue M, Thermo Fisher Scientific, USA). Pre-cleaned $SiO_2$/Si substrates were first spun-coated with 0.02 mol/L $Na_2MoO_4$ solution and then were placed at the center of the furnace. Dodecyl mercaptan solution was put in a bubbler and connected to the inlet of the system. The furnace was heated to the growth temperature of 850 °C and dodecyl mercaptan was bubbled with 10 sccm Ar into the reactor for one hour. Finally, the reactor was cooled to room temperature in Ar flow.

**Growth of $MoS_2$ by conventional CVD using solid precursors.** Growth experiments were conducted in the same CVD furnace. Sulfur powder (100 mg) was loaded in the first zone upstream, and $MoO_3$ (5−10 mg) with face-down $SiO_2$/Si substrates was placed downstream (8−10 cm) for $MoS_2$ growth. First, the tube was flushed with 350 sccm of Ar for 30 min, then heated to the growth temperature of 750 -780 °C with a ramp rate of 30 °C min$^{-1}$ and kept for

3−10 min. After growth, the furnace was cooled to room temperature. Ar (50 sccm) was introduced during the whole process. A similar method was used to grow $MoS_2$ with salt, except that 1mg KI powder was mixed with 10 mg $MoO_3$ powder and used as precursors.

**Materials characterization.** The morphology of the samples was examined by an optical microscope (Carl Zeiss Microscopy, Germany). Low magnification and high resolution TEM analyses were carried out in FEI Tecnai F30, USA with an energy-dispersive X-ray (EDX) spectroscopy detector at an acceleration voltage of 80 kV. ADF-STEM analyses were carried out using a FEI Titan Themis G2 double aberration corrected TEM. The microscope was operated at an electron acceleration voltage of 60 kV. The convergence angle was about 30 mrad and the collection range of the STEM image was around 53−200 mrad. Reflectance contrast spectra were conducted in a home-built optical measurement system. Raman and PL spectra and mappings were collected using a 532 nm laser excitation with a beam size of ~1 μm (Horiba LabRAB HR Evolution, Japan). The low temperature PL measurements were conducted using a temperature stage (Linkam HFS600E-PB4) connected to a liquid nitrogen dewar bottle.

**Density functional theory calculations.** All calculations were performed using the density functional theory as implemented in Vienna Ab-initio Simulation Package.[50] The core-electron interactions and electron exchange-correlation function were described by projected augmented wave methods [51] and Perdew-Burke-Ernzerhof generalized gradient approximation,[52] respectively. The DFT-D3 method of Grimme [53] was used to account for van der Waals force. The climbing image nudged elastic band method was used to locate the transition states to determine energy barriers.[54] The defective $MoS_2$ sheet was modelled by a 6×6×1 supercell of

MoS$_2$ with a single SV. A k-point sampling of 3×3×1 was used in the calculations. A sufficiently larger distance of 20 Å along out-of-plane direction was selected to simulate isolated monolayer properties. The kinetic energy cutoff for the plane wave basis was set to 400 eV. All the structures were fully relaxed until the force on each atom was less than 0.1 eV/Å.

## ASSOCIATED CONTENT

Supporting Information can be found with this article online.

## ACKNOWLEDGMENTS

We thank Qiangmin Yu, Zhengyang Cai, Jiaman Liu and Zenglong Guo for fruitful discussions. We acknowledge support by the NSFC (Nos. 51722206, 51991340, 51991343, 51950410577，11974156 and 51920105002), the Guangdong Innovative and Entrepreneurial Research Team Program (No. 2017ZT07C341), Guandong International Science Collaboration Project (Grant No. 2019A050510001), the Bureau of Industry and Information Technology of Shenzhen for the "2017 Graphene Manufacturing Innovation Center Project" (No. 201901171523), the Shenzhen Basic Research Project (Nos. JCYJ20190809180605522 and JCYJ20170407155608882), and also the assistance of SUSTech Core Research Facilities, especially technical support from Pico-Centre.